\documentclass[10pt]{JHEP3}

\usepackage{amsmath,amssymb,graphicx,caption}

\newcommand\fverb{\setbox\pippobox=\hbox\bgroup\verb}
\newcommand\fverbdo{\egroup\medskip\noindent%
                        \fbox{\unhbox\pippobox}\ }
\newcommand\fverbit{\egroup\item[\fbox{\unhbox\pippobox}]}
\newbox\pippobox

\title{\Large\bf Constraints of the $B_{\mu}/\mu$ Solution due to the Hidden Sector Renormalization}

\preprint{SNUTP 08-001}

\author{
Hae Young Cho  \\
  FPRD and Department of Physics and Center for Theoretical Physics,
  Seoul National University, Seoul 151-747, Korea \\
  E-mail: \email{hycho@phya.snu.ac.kr} }

\maketitle

\abstract{We investigate the validity of an idea that the $B_{\mu}$
problem is solvable via the renormalization effect in the strongly
interacting hidden sector within the gauge mediated supersymmetry
breaking scenario. Our analysis starts with a naive boundary
condition, which is that the squared scalar masses experience
$16\pi^2$ suppression. We use \texttt{softsusy} to get the low
energy spectra of superparticles with the boundary condition at the
scale ($\Lambda_{CFT}$) where the hidden sector is integrated out.
We visit the low energy spectra and return to $\Lambda_{CFT}$ where
the boundary conditions are given. We find that there is a sign
problem, which seems to be generic.}


\begin{document}

\def\lsl{ l \hspace{-0.45 em}/}
\def\ksl{ k \hspace{-0.45 em}/}
\def\qsl{ q \hspace{-0.45 em}/}
\def\psl{ p \hspace{-0.45 em}/}
\def\ppsl{ p' \hspace{-0.70 em}/}
\def\dsl{ \partial \hspace{-0.45 em}/}
\def\Dsl{ D \hspace{-0.55 em}/}
\def\matrix{ \left(\begin{array} \end{array} \right) }
\def\frsqotw{\frac{1}{\sqrt2}}
\def\frsqoth{\frac{1}{\sqrt3}}
\def\frsqtth{\sqrt{\frac{2}{3}}}
\def\frsqo{\frac{\omega}{\sqrt{3}}}
\def\frsqoo{\frac{\omega^2}{\sqrt{3}}}

\def\hf{\textstyle{\frac12~}}
\def\hff{\textstyle{\frac13~}}
\def\hfg{\textstyle{\frac23~}}
\def\DQ{$\Delta Q$}

\def\Qem{Q_{\rm em}}

\thispagestyle{empty}

\baselineskip 0.6cm

%%%%%%%%%%%%%%%%%%%%%%%%%%%%%%%%%%%%%%%%%%%%%%%%%%%%%%%%%%%%%%%%%%%%
%%%%%%%%%%%%%%%%%%%%%%%%%%%%%%%%%%%%%%%%%%%%%%%%%%%%%%%%%%%%%%%%%%%%

\section{Introduction}
Some attractive outcomes of TeV scale SUSY are a solution of the
gauge hierarchy problem \cite{Witten:1981kv}, the existence of dark
matter candidate with the R-parity \cite{Goldberg:1983nd}, and the
radiative breaking of the electroweak symmetry \cite{RSSB}. The TeV
scale soft SUSY breaking parameters are derived from the source of
SUSY breaking, presumably at the hidden sector, and the messenger
sector which couples to both the observable and the hidden sectors.
In the visible sector, the standard model (SM) degrees of freedom
are accommodated; in the messenger sector, we introduce the carriers
of the SUSY breaking information to the observable sector. Thus, the
phenomenology of the minimal supersymmetric standard model (MSSM)
depends on where the hidden sector breaks SUSY and how this
information is transmitted to the visible sector. The most popular
mediation scheme is the minimal super gravity (mSUGRA) because the
gravity can couple to both the hidden and the observable sectors. In
mSUGRA, the information is revealed as non-renormalizable operators
suppressed by powers of the Planck mass $M_{Pl}$
\cite{Nilles:1983ge}. There is, however, the serious flavor changing
neutral current (FCNC) problem in mSUGRA. The FCNC problem is
improved if one introduces very small squark mass differences
$\Delta \tilde{m_i}^2$ or very large squark masses $\tilde{m_i}^2$
\cite{Hall:1990ac}. Since TeV scale SUSY does not permit very large
$\tilde{m_i}^2$, there should be unnatural conditions such as the
universal soft breaking terms for the scalar masses. Another
scenario is the anomaly mediation scenario (AMSB), where the FCNC
amplitude can be suppressed by the scale of the distance between the
hidden sector and the visible sector branes \cite{Randall:1998uk}.
AMSB, however, has an intrinsic serious problem that the scalar
partners of the lepton are tachyonic at the low energy scale. To
remedy this problem, one has to introduce a baroque structure
\cite{modiAMSB}.

The gauge mediated SUSY breaking (GMSB) scenario was introduced as
another alternative and seems to remain as the simplest solution of
the SUSY FCNC problem \cite{GMSB}. Recently, the concept of the
metastable vacua in the hidden sectors brought renaissance of the
GMSB scenario \cite{metaGMSB}. Especially, this concept made the
GMSB scheme accommodated in the string theory
\cite{Kim:2007zj,Kawano:2007ru,GMSBst} so that the interest on GMSB
has increased. In GMSB, all SUSY breaking dimensionful parameters
can be obtained by gauge interactions except $B_{\mu}$.\\

On the other hand, $\mu$ is very strange in the MSSM because it is a
unique dimensionful parameter in the supersymmetric part of the
MSSM. So, if we consider the unbroken SUSY at a considerably high
scale, then it is natural to consider $\mu$ of that scale, for
example the order of the Planck mass $M_{Pl}$. In addition, if one
considers the $U(1)_{PQ}$ symmetry at the electroweak scale, the
$\mu$ term is forbidden. If a non-trivial Kalher potential is
considered, it is known that the SUSY breaking sector and the Higgs
fields can be coupled. Therefore, the existence of SUSY breaking can
give a rise to a correct order for $\mu$ and $B_{\mu}$ terms
\cite{Giudice:1988yz}. However, this supergravity generation of
$\mu$ assumes negligible tree level $\mu$ contribution in the
superpotential. The plausible reason for forbidding the $\mu$ term
in the superpotential is originated from symmetries such as
$U(1)_{PQ}$ and/or  $U(1)_{R}$. In general, it is difficult to
obtain the $\mu$ term in the GMSB scenario though it is easily
implemented in mSUGRA\cite{Kim:1983dt,Giudice:1988yz}, because
$U(1)_{PQ}$ cannot be broken by gauge interactions. Thus, it is
required for $U(1)_{PQ}$ breaking terms to enter the superpotential.
To generate $\mu$ and $B_{\mu}$, we introduce the direct interaction
between Higgs and messengers as follows
\begin{equation}\label{Higgsp}
\begin{split}
W_{H_{1}H_{2}}=\xi_{1}H_{1}\psi_{1}\bar{\psi_{2}}
+\xi_{2}H_{2}\bar{\psi_{1}}\psi_{2},
\end{split}
\end{equation}
where $\psi_{1}$ and $\psi_{2}$ are the messenger fields carrying
appropriate weak and hyper charges to couple to $H_{1,2}$ at the
tree level. After integrating out rather massive messengers, we get
the appropriate operators for $\mu$ and $B_{\mu}$. There is,
however, another problem in GMSB: if we use the superpotential in
(\ref{Higgsp}), $\mu$ and $B_{\mu}$ are generated at a same loop
level so that it is hard to satisfy low energy phenomenology.
Several studies on this topic \cite{muGMSB} exist already. Recently,
the role of the hidden sectors has been raised \cite{Cohen:2006qc}.
Using this idea, an alternative solution for the $B_{\mu}/\mu$
problem is suggested in the GMSB setup
\cite{Murayama:2007ge,Roy:2007nz}, and we will discuss how this
works in section 2. In this work, we investigate what conditions are
required for the $B_{\mu}/\mu$ problem in the setup of Refs.
\cite{Murayama:2007ge,Roy:2007nz}. We set the boundary condition for
this study that the squared scalar masses are $16\pi^2$ suppressed
compared to the gaugino mass squared. This is similar to the usual
gaugino mediation \cite{gMSB} in the ratio of the gaugino mass and
the scalar mass. However, the messenger scales are quite different
in the two cases so that the mass spectra at low energy may be
totally distinguishable. As a result, we find that the idea
suggested in \cite{Murayama:2007ge,Roy:2007nz} has a tachyonic
sector at low energy. We pursue the study on the region where low
energy spectra satisfy the experimental result. We trace back to the
`effective' messenger scale, where the boundary conditions, which
contain the hidden sector RG effects, are given. We find that
$B_{\mu}$ carries opposite sign to $\mu$ at the `effective'
messenger scale. If the visible sector running effects do not give a
significant contribution, this relation holds to the scale where the
operators for $B_{\mu}$ and $\mu$ are generated, and is not
compatible with the original relation between $\mu$ and $B_{\mu}$.\\

In section 2, we will briefly review on the mechanism and the menace
of tachyonic stau in the low energy spectra. In section 3, we will
obtain the low energy spectra and postulate the valid parameter
region in the sense of the low energy spectra. In section 4, we will
discuss on the consistency of this mechanism, and make a conclusion.

\section{The Basic Scheme}
One of the most important success of SUSY is that it can explain how
the electroweak symmetry breaking occurs. This can be achieved by
the stop loop. The potential for Higgs in MSSM is given as
\begin{equation} \label{hip}
\begin{split}
V=&|\mu|^2(|H_{1}|^2+|H_{2}|^2)
+\frac{1}{8}(g_1^2+g_2^2)(|H_{1}|^2-|H_{2}|^{2})^2\\
&+m^2_{H_{1}}|H_{1}|^2+m^{2}_{H_{2}}|H_{2}|^2-(B_{\mu}
H_{1}H_{2}+c.c.).
\end{split}
\end{equation}
From this we can derive the condition for EWSB, using the hessian
for the Higgs mass matrix at the origin
\begin{equation}
B_{\mu}^2>(|\mu|^2+m^2_{H_{1}})(|\mu|^2+m^2_{H_2}).
\end{equation}
Moreover, we require that the Higgs potential is bounded from below.
There is a possible dangerous direction in (\ref{hip}). Therefore,
this implies
\begin{equation}
2B_{\mu}<2|\mu|^2+m_{H_1}^2+m_{H_2}^2.
\end{equation}
For the CP even Higgs fields the mass matrix is given as
\begin{equation}
\left(
  \begin{array}{c}
    h \\
    H \\
  \end{array}
\right)=\sqrt{2} \left(
  \begin{array}{cc}
    \cos{\alpha} & -\sin{\alpha} \\
    \sin{\alpha} & \cos{\alpha} \\
  \end{array}
\right)\left(
         \begin{array}{c}
           ReH_{2}-v_2 \\
           ReH_{1}-v_1 \\
         \end{array}
       \right).
\end{equation}
With the quantum correction, the mass of the lightest Higgs field
saturates this inequality
\begin{equation}\label{higup}
m_{h}^2\lesssim
\cos^2{2\beta}M_{Z}^2+\frac{3\alpha_2}{2\pi}\frac{m_{t}^4}{M_{Z}^2}\ln{\frac{\tilde{m_{t}}^2}{M_{Z}^2}}.
\end{equation}
From this if $\tan{\beta}>4$, we can consider the lightest CP even
Higgs field as the SM Higgs field. From the result of the LEP we
know that the lower bound of the SM Higgs mass is $114.4$GeV
\cite{:2001xx}.\footnote{If the R-parity is broken, the lightest
Higgs mass can be lower than the LEP bound \cite{Rpa}.}\\

At the intermediate scale, which is between the electroweak ($M_Z$)
and the messenger ($M_{mess}$) scales, the MSSM couplings are not so
large. Therefore, once $B_{\mu}$ and $\mu$ are generated, the ratio
between $B_{\mu}$ and $\mu^2$ does not suffer a significant change.
This is undesirable at the electroweak scale. For successful
electroweak symmetry breaking, we require that both of these are
order of the gaugino masses,
\begin{equation}\label{phereq}
\begin{split}
B_{\mu}\sim \mu^2 \ \ \ \ \mu\sim m_{\frac{1}{2}}.
\end{split}
\end{equation}
Let us consider how the hidden sector strong dynamics works toward
the electroweak symmetry breaking. For a concrete discussion of the
messenger effect toward the MSSM physics and the $\mu$ generation,
we adopt the simple superpotential in (\ref{Higgsp}). Integrating
out the heavy messenger fields ($M_{mess}\gg \Lambda_{CFT}\gg M_Z$),
we obtain non-renormalizable interaction terms between the hidden
and the visible sector fields. In this way, let us consider the
following operators relevant for the dimensionful parameters in the
MSSM,
\begin{equation}\label{dimfpara}
\begin{split}
&O_{\phi}: \int d^4 \theta~
c_{\phi}^{q}~\frac{q^{\dagger}q}{M^2}\phi^{\dagger}\phi, \ \ \int
d^4\theta~
c_{\phi}^{s}~\frac{S^{\dagger}S}{M^2}\phi^{\dagger}\phi,\\
&O_{B_{\mu}}: \int d^4 \theta~c_{B_{\mu}}^{q}~
\frac{q^{\dagger}q}{M^2}H_{1}H_{2}+h.c., \ \ \int d^4
\theta~ c_{B_{\mu}}^{s}~\frac{S^{\dagger}S}{M^2}H_{1}H_{2}+h.c., \\
&O_{\lambda}: \int d^4\theta~
c_{\lambda}^{s}~\frac{S}{M}W^{a\alpha}W^{a}_{\alpha}+h.c,\\
&O_{A}: \int d^4\theta~
c_{A}^{s}~\frac{S}{M}\phi^{\dagger}\phi+h.c.,\\
&O_{\mu}: \int d^4\theta~
c_{\mu}^{s}~\frac{S^{\dagger}}{M}H_{1}H_{2}+h.c.,
\end{split}
\end{equation}
where $H_{1,2}$ and $\phi$ are the MSSM fields and the rest are the
intermediate scale fields which constitute the ingredients for SUSY
breaking. Here, $S, q$ and $W^\alpha$ are spurion, quark, and
gaugino fields, respectively, in the intermediate scale, and $c$s
are the couplings. Refs. \cite{Murayama:2007ge,Roy:2007nz} consider
a hidden conformal sector at the intermediate scale, which is
guaranteed by Seiberg's duality \cite{Seiberg:1994pq}. In Seiberg's
conformal window, the electric and magnetic descriptions are the
same. At this window the gauge coupling is asymptotically free, and
hence it is meaningless to use the perturbation method in the low
energy limit. Thus, the theory naturally has a low energy cutoff,
which is usually represented as a mass parameter $\Lambda$. In QCD,
for example, it is denoted as $\Lambda_{QCD}$. In this vein, we will
define the intermediate mass scale as $\Lambda_{CFT}$. Next,
integrating out the fields at $\Lambda_{CFT}$, the effective
operators for the soft terms are obtained. That is to say, the
renormalization effect below the scale $\Lambda_{CFT}$ is nothing
but that of MSSM with the boundary conditions fixed at
$\Lambda_{CFT}$. The MSSM RG has been widely studied, in the
literature such as \cite{Kazakov:2000us}.

From (\ref{dimfpara}) we note that the soft scalar mass has the same
property as $B_{\mu}$. On the other hand, the trilinear coupling $A$
behaves the same as the gaugino mass or $\mu$. From (\ref{Higgsp})
we also note that $\mu$ and the gaugino mass are generated at one
loop level. So the relative size between the gaugino mass and $\mu$
can be easily fitted to the phenomenological expectation. However,
not only $\mu$ but also $B_{\mu}$ are generated at one loop level.
It turns out that the ratio between $B_{\mu}$ and $\mu$ at the
messenger scale is too large to fit the phenomenological
requirement. The renormalization in the SUSY gauge theory is
revealed as the wave function renormalization. Considering the 1PI
renormalization for $S^{\dagger}S$ in addition to the wave function
renormalization below $\Lambda_{CFT}$, then the effective operators
in (\ref{dimfpara}) should be substituted by
\begin{equation}\label{consider1pi}
\begin{split}
&O_{\phi}: \int d^4 \theta
\left(\frac{\Lambda_{CFT}}{M_{mess}}\right)^{\alpha_q}Z_q^{-1}
~c_{\phi}^{q}~\frac{q^{\dagger}q}{M^2}\phi^{\dagger}\phi,
\\ &\quad\quad\int d^4\theta \left(\frac{\Lambda_{CFT}}{M_{mess}}\right)^{\alpha_s}Z_s^{-1}~
c_{\phi}^{s}~\frac{S^{\dagger}S}{M^2}\phi^{\dagger}\phi, \\
&O_{B_{\mu}}: \int d^4 \theta
\left(\frac{\Lambda_{CFT}}{M_{mess}}\right)^{\alpha_q}Z_q^{-1}~
c_{B_{\mu}}^{q}~\frac{q^{\dagger}q}{M^2}H_{1}H_{2}+h.c.,
\\ &\quad\quad\int d^4
\theta
\left(\frac{\Lambda_{CFT}}{M_{mess}}\right)^{\alpha_s}Z_s^{-1}~
c_{B_{\mu}}^{s}~\frac{S^{\dagger}S}{M^2}H_{1}H_{2}+h.c., \\
&O_{\lambda}: \int d^4\theta Z_s^{-1/2}~
c_{\lambda}^{s}~\frac{S}{M}W^{a\alpha}W^{a}_{\alpha}+h.c,\\
&O_{A}: \int d^4\theta Z_s^{-1/2}~
c_{A}^{s}~\frac{S}{M}\phi^{\dagger}\phi+h.c.,\\
&O_{\mu}: \int d^4\theta
Z_s^{-1/2}~c_{\mu}^{s}~\frac{S^{\dagger}}{M}H_{1}H_{2}+h.c.,
\end{split}
\end{equation}
where $\alpha_q$ and $\alpha_s$ are anomalous dimensions and the
wave function renormalization factors $Z_s$ and $Z_q$ are defined as
\begin{equation}
Z_{s,q}=\left(\frac{\Lambda_{CFT}}{M_{mess}}\right)^{3R(S,q)-2}.
\end{equation}
There exist some subtle points in this mechanism which are caused by
the mixing between operators given in (\ref{consider1pi}). There can
be mixing between the quadratic operator and the linear operators as
well as mixing between the quadratic operators. Each mixing is
induced by direct interactions between matter and messenger fields,
and the strongly interacting hidden sector respectively. The latter
appears as the anomalous dimension, although we can not get the
exact value. Here we assume that we obtain the hidden sector effect
as
\begin{equation}\label{deal}
\text{suppression
factor}=\left(\frac{\Lambda_{CFT}}{M_{mess}}\right)^{\alpha},
\end{equation}
where $\alpha$ is the smallest eigenvalue of matrix for the
anomalous dimensions, which reflects the mixing between the
quadratic operators. And the other affects the boundary conditions
too. In ref. \cite{Giudice:1997ni}, the effects on the soft
parameters under the presence of such a superpotential
(\ref{Higgsp}) are shown. Now we will turn to very subtle points of
mixing between operators. In (\ref{consider1pi}), we do not consider
the operator mixing, however, within the sense of effective field
theory there is mixing, which affects on the boundary condition
which will be shown in the next section. Generally, the soft
parameters of mass dimension two are some combinations of the
quadratic operators and the linear operators. We, therefore, provide
the terms which appear as the boundary conditions. This is well
explained in ref.\cite{Murayama:2007ge}, and we follow its
description. For the ordinary scalar fields, it is
$c_{\phi}^{s}-|c_A^{s}|^2$. For the Higgs field, there is another
contribution from $\mu$ so that it is
$c_{H_{1,2}}^s-|c_A^{s}|^2-|c_{\mu}^s|^2$. Finally, it is
$c_{B_{\mu}}^{s}-c_{\mu}^s(c_{A_{H_1}}^s+c_{A_{H_2}}^s)$ for
$B_{\mu}$. These all experience the hidden sector RG effect, and it
should be realized in the boundary conditions. Via these effects at
the scale of CFT breaking, the ratio of $B_{\mu}$ and $\mu$ can be
made to satisfy the relation (\ref{phereq}). Then we can see that
$B_{\mu}$ suffers the renormalization effects through the strongly
interacting CFT sector. However, there is an effect which we should
not ignore. It is that the squared masses of scalar also suffer the
same kind of renormalization as $B_{\mu}$. In the operator sense,
scalar masses and gaugino masses are generated properly at the the
messenger scale. However, when we reach at the scale
$\Lambda_{CFT}$, the scalar masses would suffer $16\pi^2$
suppression by the same mechanism that reduces $B_{\mu}$. On the
other hand, the trilinear
$A$ terms do not undergo such a suppression.\\
Keeping these in mind, let us consider the mixing matrices for the
$\tilde{m}^{2}_t,\tilde{m}^{2}_b$ and $\tilde{m}^{2}_\tau$:
\begin{equation} \label{stopmat}
\left(\begin{array}{cc} \tilde m_{tL}^2& m_t(A_t-\mu\cot \beta )
\\ m_t(A_t-\mu\cot \beta ) & \tilde m_{tR}^2 \end{array}  \right),
\end{equation}
\begin{equation} \label{sbottommat}
\left(\begin{array}{cc} \tilde  m_{bL}^2& m_b(A_b-\mu\tan \beta )
\\ m_b(A_b-\mu\tan \beta ) & \tilde  m_{bR}^2 \end{array}
\right),
\end{equation}
\begin{equation} \label{staumat} \left(\begin{array}{cc}
\tilde  m_{\tau L}^2& m_{\tau}(A_{\tau}-\mu\tan \beta ) \\
m_{\tau}(A_{\tau}-\mu\tan \beta ) & \tilde m_{\tau R}^2
\end{array}  \right)
\end{equation}
with
\begin{equation*}
\begin{split}
&\tilde m_{tL}^2=\tilde{m}_Q^2+m_t^2+\frac{1}{6}(4M_W^2-M_Z^2)\cos
  2\beta ,\\
&\tilde m_{tR}^2=\tilde{m}_U^2+m_t^2-\frac{2}{3}(M_W^2-M_Z^2)\cos
  2\beta ,\\
&\tilde m_{bL}^2=\tilde{m}_Q^2+m_b^2-\frac{1}{6}(2M_W^2+M_Z^2)\cos
  2\beta ,\\
&\tilde m_{bR}^2=\tilde{m}_D^2+m_b^2+\frac{1}{3}(M_W^2-M_Z^2)\cos
  2\beta ,\\
&\tilde m_{\tau
L}^2=\tilde{m}_L^2+m_{\tau}^2-\frac{1}{2}(2M_W^2-M_Z^2)\cos
 2\beta ,\\
&\tilde m_{\tau R}^2=\tilde{m}_E^2+m_{\tau}^2+(M_W^2-M_Z^2)\cos
  2\beta .
\end{split}
\end{equation*}
Now, we can obtain the masses at the electroweak scale. As denoted
above, below $\Lambda_{CFT}$ the renormalization equations are that
of the MSSM. Thus, the renormalization property of each
diemensionful parameter is given as
\begin{equation*}\label{MSSMrg1}
\begin{split}
\frac{dM_i}{dt}  =&  b_i \alpha_iM_i . \\
\frac{dA_U}{dt}  =& \frac{16}{3}\alpha_3 M_3 + 3\alpha_2M_2
+ \frac{13}{15}\alpha_1 M_1+6Y_UA_U+Y_DA_D,\\
\frac{dA_D}{dt}  =& \frac{16}{3}\alpha_3 M_3 + 3\alpha_2 M_2
+ \frac{7}{15}\alpha_1 M_1+6Y_DA_D+Y_UA_U+Y_LA_L,  \\
\frac{dA_L}{dt}  =&  3\alpha_2 M_2 + \frac{9}{5}\alpha_1
M_1+3Y_DA_D+4Y_LA_L,\\
\frac{dB}{dt}  =&  3\alpha_2 M_2 + \frac{3}{5}\alpha_1
M_1+3Y_UA_U+3Y_DA_D+Y_LA_L,
\end{split}
\end{equation*}
\begin{equation*}
\begin{split}
\frac{d\tilde{m}^2_Q}{dt}=&-\left[ (\frac{16}{3}\alpha_3M^2_3 +
3\alpha_2M^2_2 + \frac{1}{15}\alpha_1M^2_1)
-Y_U(\tilde{m}^2_Q+\tilde{m}^2_U+m^2_{H_2}+A^2_U)\right. \\
&\left. -Y_D(\tilde{m}^2_Q+\tilde{m}^2_D+m^2_{H_1}+A^2_D)\right],\\
\frac{d\tilde{m}^2_U}{dt} =& - \left[(\frac{16}{3}{\alpha}_3M^2_3
+\frac{16}{15}{\alpha}_1M^2_1)-2Y_U(\tilde{m}^2_Q+\tilde{m}^2_U+m^2_{H_2}+A^2_U)\right] ,\\
\frac{d\tilde{m}^2_D}{dt} =&
 - \left[(\frac{16}{3}{\alpha}_3M^2_3+ \frac{4}{15}{\alpha}_1M^2_1)
-2Y_D(\tilde{m}^2_Q+\tilde{m}^2_D+m^2_{H_1}+A^2_D)\right],\\
\frac{d\tilde{m}^2_L}{dt}  =& -\left[3(
 {\alpha}_2M^2_2 + \frac{1}{5}{\alpha}_1M^2_1)
-Y_L(\tilde{m}^2_L+\tilde{m}^2_E+m^2_{H_1}+A^2_L)\right],\\
\frac{d\tilde{m}^2_E}{dt} =& -\left[ (
 \frac{12}{5}\alpha_1M^2_1)-2Y_L(
\tilde{m}^2_L+\tilde{m}^2_E+m^2_{H_1}+A^2_L)\right],
\end{split}
\end{equation*}
\begin{equation*}\label{MSSMrg2}
\begin{split}
\frac{d\mu^2}{dt}=& -\mu^2\left[3(\alpha_2+
\frac{1}{5}\alpha_1)-(3Y_U+3Y_D+Y_L)\right], \\
\frac{dm^2_{H_1}}{dt}=& -\left[3({\alpha}_2M^2_2 +\frac{1}{5}{
\alpha}_1M^2_1)
-3Y_D(\tilde{m}^2_Q+\tilde{m}^2_D+m^2_{H_1}+A^2_D)\right.\\
& \left. -Y_L(\tilde{m}^2_L+\tilde{m}^2_E+m^2_{H_1}+A^2_L)\right],\\
\frac{dm^2_{H_2}}{dt}  =& -\left[ 3({\alpha}_2M^2_2
+\frac{1}{5}{\alpha}_1M^2_1)
-3Y_U(\tilde{m}^2_Q+\tilde{m}^2_U+m^2_{H_2}+A^2_U)\right],
\end{split}
\end{equation*}
where $\alpha_{i}=\frac{g_i^2}{4\pi}$ and
$t=\ln{\frac{Q}{\mu_{r}}}$. Here, we observe that the right-handed
stau can be dangerous. Because it carries only $U(1)$ hypercharge it
receives the contribution from bino mass. Thus, the gaugino mass
contribution cannot be large via the MSSM renormalization. On the
other hand, $A_{\tau}$ can not be neglected in general at the
$\Lambda_{CFT}$ scale. From the mixing matrix for the stau
(\ref{staumat}), we can expect that stau becomes tachyonic ina
certain range of parameters. To show it explicitly, we use
\texttt{softsusy} for the scalar masses running
\cite{Allanach:2001kg}. Even if the trilinear couplings are made
zero at the messenger scale, the scalar masses are not free from the
menace of the tachyonic states, because there are also contributions
from $\mu$ as can be seen from
(\ref{stopmat},\ref{sbottommat},\ref{staumat}). We should be careful
about this effect too.

\section{Numerical Analysis : Visiting the Low Energy Spectra}

To check the discussion in the previous section, we use
\texttt{softsusy}. In \texttt{softsusy}, the input parameters are
$\tan{\beta}$, $M_{mess}$, number of messengers, gravity
contribution and $\Lambda=\frac{F}{M_{mess}}$. We will use
$\Lambda_{CFT}$ as an `effective' messenger scale. At this scale,
the squared scalar masses suffer $16\pi^2$ suppression and the
trilinear term can be obtained. It is a good approximation to use
the basic setup provided by \texttt{softsusy} except the suppression
of scalar masses and non-zero trilinear coupling. The others such as
$\mu$ and $B_{\mu}$ are obtained in the range where the low energy
phenomenology allows. In the previous section, we discuss that the
mixing can exist, and we will apply the argument to the boundary
conditions. Since we consider the minimal case, i.e. all MSSM fields
do not have Yukawa interaction with the messenger fields except
Higgs, there is no significant contribution to $A_{\phi}$. On the
other hand, the soft masses of the Higgs obtain these contributions
of $\delta m_{H_{1,2}}^2 \sim -\mu^2$. In addition, there is a
contribution of order of $-\mu$ to the trilinear coupling
$A_{U,D,L}$ respectively. Therefore, we set the universal trilinear
coupling $A=-\mu$. The relation between trilinear couplings is as
follows
\begin{equation*}
\begin{split}
&A_{U,D,L}=Y_{U,D,L}A, \\
&A=A_{H_{1,2}},
\end{split}
\end{equation*}
where $Y_{U,D,L}$ are Yukawa matrices.

We will do our calculations as follows:
\begin{itemize}
\item Set the scale, where the hidden sectors are integrated out, as
$\Lambda_{CFT}$.
\item Set the  messenger scale as $10^{14}$GeV.
\item At $\Lambda_{CFT}$, the scalar masses are suppressed by $16\pi^2$.
\item Set $\Lambda_{CFT}$, i.e. the `effective' messenger scale, as $10^{8}$GeV.
\item The sign of $\mu$ is positive.
\item Set the gravity contribution as zero.
\item Set $m_{t}=170.9$GeV.
\item The trilinear couplings of Higgs are generated as $\delta
A_{H_{1,2}}\sim -\mu$. By them, the universal trilinear coupling
satisfies $A\sim-\mu$.
\item The soft Higgs masses receive the contribution of $\delta
m^2_{H_{1,2}} \sim -\mu^2$.
\item Set $\tan{\beta}$ and $\frac{F}{M_{mess}}$ as free parameters.
\item Scan $\tan{\beta}$ from 4 to 50 and $\frac{F}{M_{mess}}$ from $5.0\times10^4$GeV to $2.0\times10^5$GeV for the case of 1 messenger.
\item No consideration about other low energy constraints such as $B_s\gamma$.
\end{itemize}

\begin{figure}[t]
\begin{center}
\includegraphics[width=12cm]{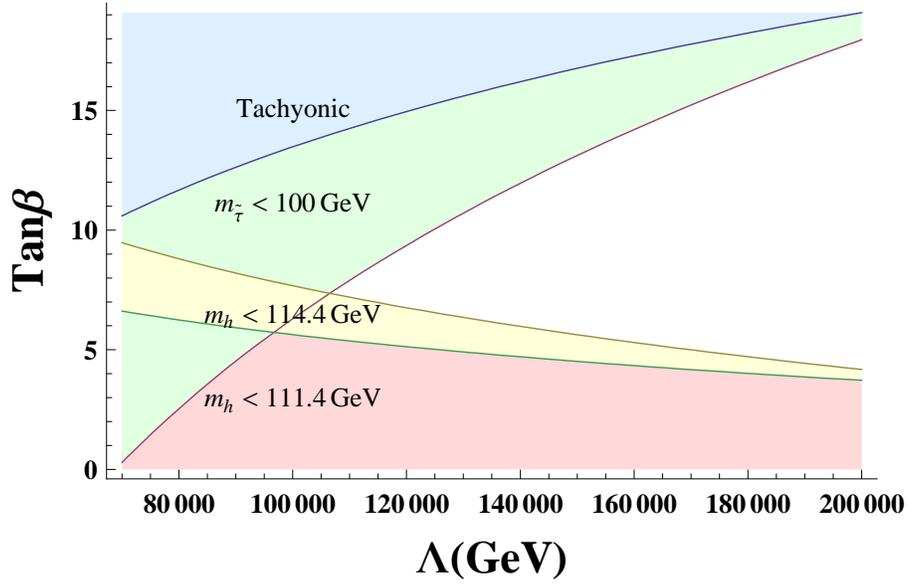}
\end{center}
\caption{ \footnotesize Plot of forbidden region in $16\pi^2$
suppression case, with $A=-\mu$, $\Lambda_{CFT}=10^8$GeV and the
number of messengers $=1$. The yellow and the red parts are mass
bound for the lightest Higgs.}\label{CzA}
\end{figure}

In Fig.~(\ref{CzA}) the blue section represents the tachyonic region
and the green part does the stau direct search bound. Since there
can be theoretical errors in calculating the mass spectra with
software packages such as \texttt{FeynHiggs} and \texttt{softsusy}
\cite{Allanach:2004rh}, we allow $-3$GeV difference. The yellow
region represents the section where the lightest Higgs mass is
between 111.4GeV and 114.4GeV. On the other hands, the red region
can be said to be definitely ruled out by the direct Higgs search
bound of the LEP experiment.
In this figure, we can see that there is a tachyonic region at the
large $\tan{\beta}$. This feature appears similarly in some
parameter space in the gaugino mediation. This can be easily
understood when we consider the mass matrix of the stau in
(\ref{staumat}). To use the hidden sector strong RG effects as a
solution of $B_{\mu}$ problem, the squared scalar masses are not
free from $16\pi^2$ suppression. Therefore, the diagonal parts of
the stau mass matrix are rather small compared to the off diagonal
parts. Here we look more carefully the off diagonal parts. The off
diagonal parts are composed with tau mass, the trilinear coupling
$A$ and $\mu\tan{\beta}$. In \texttt{softsusy}, $\mu$ and $B_{\mu}$
are fitted by the proper EWSB; therefore, we do not have to worry
about this. We investigate the possibility that the parameter space
can be enlarged. Let us consider the case that the `effective'
messenger scale to be different from $10^8$GeV. Varying
$\Lambda_{CFT}$ from $10^{6}$GeV to $10^{10}$GeV, we find that the
pattern does not change significantly. As denoted above we have the
messenger scale as $10^{14}$GeV; thus, the `effective' messenger
scale can not be larger than it.

\section{Numerical Analysis : More on the Valid Region}

The mechanism which we investigate, also provides generating $\mu$
on a theoretical base. Here we will check whether the relation in
(\ref{phereq}) can be satisfied at the scale where $16\pi^2$
suppression does appear. We will stay in the region where the low
energy spectra appear to be valid. First of all, we should keep in
mind that in \texttt{softsusy} $\mu$ and $B_{\mu}$ are fitted by the
requirement of the proper EWSB. The method to consider the hidden
sector RG running effects has a unique property. Since the hidden
sector RG effects which make the $B_{\mu}$ comparable to $\mu^2$,
affect the operators which are universally proportional to
$SS^{\dagger}$; therefore, the squared scalar masses suffer such a
suppression. These effects are revealed in the boundary conditions
which we have chosen at the `effective' messenger scale. Let us
return to the start point of our analysis. We consider a direct
interaction between messengers and Higgs like (\ref{Higgsp}). With
this superpotential, we derive these relations:
\begin{equation} \label{or}
\begin{split}
&\mu=\frac{\xi_{1}\xi_{2}}{16\pi^2}\Lambda
f(\lambda_1/\lambda_2)\left[1+\mathcal{O}(\frac{F^2}{M_{mess}^4})\right]
\\
&B_{\mu}=\frac{\xi_{1}\xi_{2}}{16\pi^2}\Lambda^2
f(\lambda_1/\lambda_2)\left[1+\mathcal{O}(\frac{F^2}{M_{mess}^4})\right]
\\
&B_{\mu}=\Lambda\mu,
\end{split}
\end{equation}
where $\lambda_{1,2}$ are coupling constants between the messenger
and the goldstino supermultiplet. $f$ is a function appearing after
we integrate out the messenger fields. To satisfy phenomenological
low energy requirements, we introduce hidden sector RG effects. As a
result we get the suppression factor appear in (\ref{consider1pi}).
We set the boundary conditions to represent such factors and we get
the valid region. Now let us refer to
$\tilde{B_{\mu}}$\footnote{From now on, the tilded represent the
value which obtained by \texttt{softsusy} to satisfy the
requirements of the low energy} as the postulated value by
\texttt{softsusy} to satisfy the low energy requirements and
$B_{\mu}^{\prime}$\footnote{From now on, the primed are obtained by
the trace back RG. For example, $\mu^{\prime}$ is the result of the
trace back RG of $\tilde{\mu}$.} as one obtained by the trace back
RG of $\tilde{B_{\mu}}$ respectively. The ratio
$\frac{B_{\mu}^{\prime}}{B_{\mu}}$, which affects the squared scalar
masses, is set as boundary conditions at the `effective' messenger
scale. Then we expect that the region which passes our consistency
test, would appear as a band in the valid region of section 3. The
band should be under the control of the value $\xi_{1}\xi_{2}$.
However, we should not miss a point that $\mu$ is also dependant on
the value of $\xi_{1}\xi_{2}$. If we use the third relation of
(\ref{or}), we can eliminate this
$\xi_{1}\xi_{2}$ dependance.\\

\begin{figure}[t]
\begin{center}
\includegraphics[width=12cm]{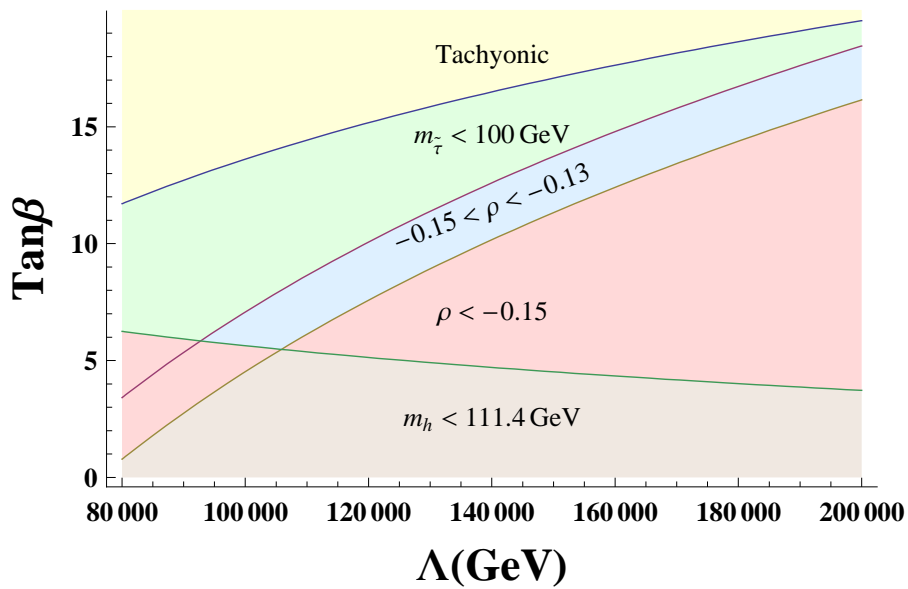}
\end{center}
\caption{\footnotesize The set up is the same as the previous one
except that the yellow part is excluded and we provide the ratio
$\rho$ between the theoretical prediction and the result of the
trace-back RG. The blue region is $-0.15<\rho<-0.13$, and the red is
$\rho<-0.15$.}\label{CzA2}
\end{figure}

\begin{figure}[t]
\begin{minipage}{68mm}
\includegraphics[width=0.9\textwidth]{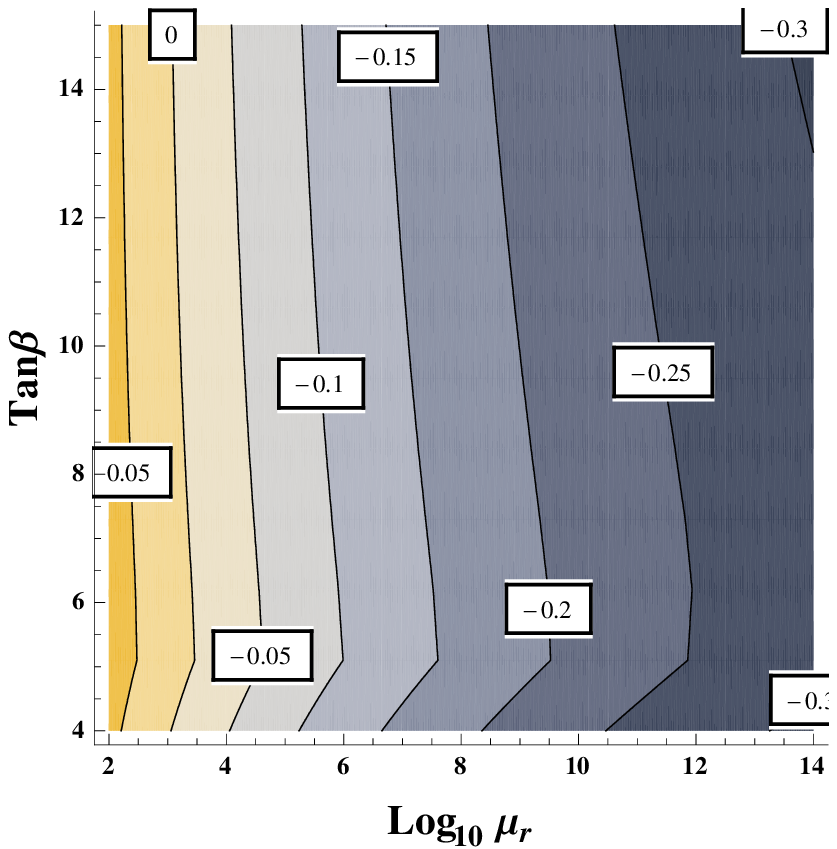}
\caption{\footnotesize The RG of $B_{\mu}^{\prime}$ with
$\Lambda=1.5\times 10^5GeV$ within the visible sector only. Here we
see that the visible sector cannot make
$\frac{B_{\mu}^{\prime}}{\mu^2}$ positive.}\label{CzA9}
\end{minipage}
\hfil
\begin{minipage}{68mm}
\includegraphics[width=0.9\textwidth]{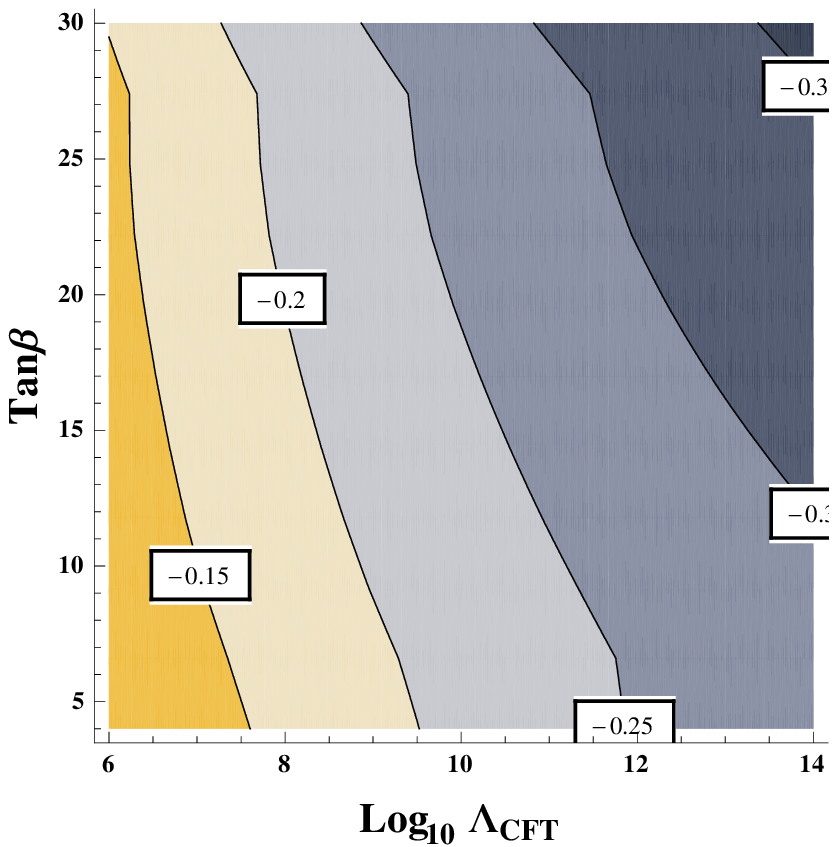}
\caption{ \footnotesize The effect of varying $\Lambda_{CFT}$ scale
with $\Lambda=1.5\times 10^5GeV$. Here we can see that the sign
problem is generic in this mechanism. The contours represent the
value of $\frac{B_{\mu}^{\prime}}{\mu^2}$. }\label{CzA4}
\end{minipage}
\end{figure}

Our strategy is very simple. Once we get the $\tilde{\mu}$ and
$\tilde{B_{\mu}}$ at the electroweak scale. We will follow the MSSM
RG flow to the `effective' messenger scale in the valid parameter
space so that we can get $\mu^{\prime}$ and $B_{\mu}^{\prime}$.
Moreover, it is natural to identify $\mu$ with $\mu^{\prime}$. Then
we will check whether the factor we get by the trace back RG is the
same as we set as boundary conditions and the evaluation of the
$\mu$ is consistent with (\ref{or}). Now let us turn to the
Fig.~(\ref{CzA2}).
In this case, we provide the ratio $\delta$ between $\mu \times
\Lambda$ and $B_{\mu}^{\prime}\times \text{suppression factor}$
obtained by the \texttt{softsusy}
\begin{equation}
\rho=\frac{B_{\mu}^{\prime}\times(\text{suppression
factor})}{\mu\times\Lambda}.
\end{equation}\label{dod}
Here we see that the valid region in Fig.~(\ref{CzA}) has a negative
$\rho$ in Fig.~(\ref{CzA2}). This means that $B_{\mu}^{\prime}$ have
an opposite sign to $\mu$. Here we want to look into (\ref{or})
carefully. The third relation says that if we fix $\mu$ to be
positive real, then the sign of $B_{\mu}$ is dependent on the sign
of $\Lambda$. That is, if $\Lambda$ is positive, then
$B_{\mu}^{\prime}$ as well as $B_{\mu}$ should be positive, since
the suppression factor does not change the sign of
$B_{\mu}^{\prime}$. If the suppression factor change the sign, it
will affect the sign of the squared scalar masses. Of course, there
are studies on this case, i.e. the negative squared scalar masses by
allowing a large mixing \cite{negsclar}, but we will leave this
topic to the further study\footnote{We run the program for the
negative suppression, and we find that in this case this mechanism
pass the our consistency test. We, however, need more
clarification.}. This sign problem might be accidental at
$\Lambda_{CFT}$, so we are not sure whether this can be really
problematic. The visible sector contribution is suppressed as much
as $B_{\mu}$ above $\Lambda_{CFT}$. Therefore, the dominant
contribution comes from the hidden sector, and it is dependent on
the sign of $B^{\prime}_{\mu}$.  To see explicitly, let us check
this. Terms, which run for $B_{\mu}$ RG in the visible sector, are
the linear terms shown in (\ref{consider1pi}); they experience the
same hidden sector RG effect as $\mu$. We can divide the RG property
of $B_{\mu}^{\prime}$ into two part, the visible sector contribution
and the hidden sector contribution:
\begin{equation*}
\begin{split}
\delta B_{\mu}^{\prime}&=\delta (\text{visible part}) +\delta
(\text{hidden contribution}) \\
&=\delta(B\times \mu \times \text{suppression factor})\\
&=\delta (B\times \mu)\times (\text{suppression factor})+
B\times\mu\times \delta(\text{suppression factor}),
\end{split}
\end{equation*}
where $B=\frac{B_{\mu}}{\mu}$, and the suppression factor is
$\left(\frac{\mu_R}{M_{mess}}\right)^{\alpha}$ above
$\Lambda_{CFT}$. As denoted above, the hidden sector contribution
depends on the sign of $B_{\mu}^{\prime}$; thus, we should check
whether the visible sector contribution can flip the sign. Here we
assume that the visible sector enjoys the ordinary MSSM RG. It is
sufficient to check the RG of $B_{\mu}^{\prime}$ with the ordinary
MSSM RG equation to the real messenger scale. Let us turn off the
hidden sector contribution for a while, i.e. $\delta (\text{hidden
sector})=0$. Then we run the visible sector RG of $B_{\mu}^{\prime}$
to the messenger scale. As a result, we find that $B_{\mu}^{\prime}$
does not get positive with the visible sector contribution only (See
Fig.~\ref{CzA9}).
We may doubt whether this sign problem is originated from the rather
small effective messenger scale. In Fig.~(\ref{CzA4}), we see that
this problem is generic; especially,
$\frac{B_{\mu}^{\prime}}{\mu^2}$ becomes smaller as $\tan\beta$
increases. On the other hand, this result is highly dependent on the
boundary condition, especially on the trilinear coupling $A$. If we
accept the minimal Yukawa coupling of the messenger fields, the
dominant contribution to $A$ is derived by the Higgs-messenger
Yukawa coupling, and this is the option we choose. If we choose $A$
differently, for example to be $\mu$, then the result change
seriously though this is not the ordinary case.
\section{Conclusion}

In the present study, we investigate the low energy spectra of the
$B_{\mu}/\mu$ solution provided by the strong hidden sector. Via the
strong hidden sector RG effects, the squared scalar masses suffer
$16\pi^2$ suppression. As a result, diagonal parts of the mass
matrices of the scalar can be relatively small compared with other
cases. Especially stau might give a constraint in the parameters
space. Using \texttt{softsusy}, we observe that there exists
tachyonic sector for the large $\tan{\beta}$. In the region which
appears to be valid in the low energy spectra test, we trace back to
the `effective' messenger scale along the MSSM RG flow. Then we
compare the factor which we obtain by the trace back RG and the
factor which we have chosen as the boundary conditions at the
`effective' messenger scale. During this study, we find that there
is a sign problem, which seems generic in this mechanism.

%%%%%%%%%%%%%%%%%%%%%%%%%%%%%%%%%%%%%%%%%%%%%%%%%%%%%%%%%%%%%%%%%%%%%%%%%%
%%%%%%%%%%%%%%%%%%%%%%%%%%%%%%%%%%%%%%%%%%%%%%%%%%%%%%%%%%%%%%%%%%%%%%%%%%

\begin{acknowledgments}

We thank Jihn E. Kim, H. D. Kim and K. J. Bae for useful discussions
and J. H. Kim for technical helps. This work was supported in part
by the Korea Research Foundation Grant funded by the Korean
Goverment(MOEHRD) (KRF-2005-084-C00001), and by the Center for
Quantum Spacetime (CQUeST), Sogang University (the KOSEF grant
R11-2005-021).

\end{acknowledgments}

%%%%%%%%%%%%%%%%%%%%%%%%%%%%%%%%%%%%%%%%%%%%%%%%%%%%%%%%%%%%%%%%%%%%%%%%%%
%%%%%%%%%%%%%%%%%%%%%%%%%%%%%%%%%%%%%%%%%%%%%%%%%%%%%%%%%%%%%%%%%%%%%%%%%%

\end{document}